\documentclass{aastex}

\begin{document}
%\setstcolor{red}

\title{ Dynamics of  descending knots in a solar prominence and their possible contributions to  the heating of the local corona}

\author{Yi Bi}
\affil{ Yunnan Observatories, Chinese Academy of Sciences, 396 Yangfangwang, Guandu District, Kunming, 650216, People's Republic of China}
\affil{Key Laboratory of Solar Activity, National Astronomical Observatories of Chinese Academy of Science, Beijing 100012, People's Republic of China}
\affil{Center for Astronomical Mega-Science, Chinese Academy of Sciences, 20A Datun Road, Chaoyang District, Beijing, 100012, People's Republic of China}

\author{Bo Yang}
\affil{ Yunnan Observatories, Chinese Academy of Sciences, 396 Yangfangwang, Guandu District, Kunming, 650216, People's Republic of China}
\affil{Center for Astronomical Mega-Science, Chinese Academy of Sciences, 20A Datun Road, Chaoyang District, Beijing, 100012, People's Republic of China}

\author{Ting Li}
\affil{Key Laboratory of Solar Activity, National Astronomical Observatories of Chinese Academy of Science, Beijing 100012, People's Republic of China}
\affil{School of Astronomy and Space Science, University of Chinese
Academy of Sciences, Beijing 100049, China}

\author{Yan Dong}
\affil{ Yunnan Observatories, Chinese Academy of Sciences, 396 Yangfangwang, Guandu District, Kunming, 650216, People's Republic of China}
\affil{Center for Astronomical Mega-Science, Chinese Academy of Sciences, 20A Datun Road, Chaoyang District, Beijing, 100012, People's Republic of China}

\author{Kaifan Ji}
\affil{ Yunnan Observatories, Chinese Academy of Sciences, 396 Yangfangwang, Guandu District, Kunming, 650216, People's Republic of China}
\affil{Center for Astronomical Mega-Science, Chinese Academy of Sciences, 20A Datun Road, Chaoyang District, Beijing, 100012, People's Republic of China}

%% Note that the \and command from previous versions of AASTeX is now
%% depreciated in this version as it is no longer necessary. AASTeX 
%% automatically takes care of all commas and "and"s between authors names.

%% AASTeX 6.2 has the new \collaboration and \nocollaboration commands to
%% provide the collaboration status of a group of authors. These commands 
%% can be used either before or after the list of corresponding authors. The
%% argument for \collaboration is the collaboration identifier. Authors are
%% encouraged to surround collaboration identifiers with ()s. The 
%% \nocollaboration command takes no argument and exists to indicate that
%% the nearby authors are not part of surrounding collaborations.

%% Mark off the abstract in the ``abstract'' environment. 
\begin{abstract}
The knots  in solar prominences are often observed   to fall with nearly constant velocity,  but the associated   physical mechanism  is currently not well understood. 
 In this letter, we presented a prominence observed by  New Vacuum Solar Telescope (NVST) in $H_{\alpha}$ wavelength.
Knots that rose within the prominence appear to have been preferentially located at higher altitude, whereas those that fell were found throughout the entire prominence structure.
The descending speed of the knots near the solar surface  was higher than that far away from the solar surface. 
We noted that the knots near the solar surface may run along a set of coronal loops observed  from  the Atmospheric Imaging Assembly.
 Elsewhere, the majority of knots are interpreted to have descended across more horizontal magnetic field with a nearly constant speed.
 This lack of acceleration indicates that the liberated gravitational potential energy may not manifest as an increase in kinetic energy.
 Assuming instead that the descending knots were capable of exciting Alfv{\'e}n waves that could then dissipate within the local corona, the gravitational potential energy of the knots may have been converted into thermal energy.
 Assuming a perfectly elastic system, we therefore estimate that the gravitational energy loss rate of these observed knots amounts to $\approx$~1/2000 of that required to heat the entire quiet-Sun, increasing to 1/320 when considering possibly further downward motions of the knots having disappeared in the $H_{\alpha}$ observations. This result suggests such a mechanism may contribute to the heating of the corona local to these prominences.
\end{abstract}
%Comparing with the images from  the Atmospheric Imaging Assembly on board the Solar DynamicObservatory (SDO), 
%% Keywords should appear after the \end{abstract} command. 
%% See the online documentation for the full list of available subject
%% keywords and the rules for their use.
%\keywords{Sun: corona - Sun: filaments, prominences - Sun: magnetic fields}

%% From the front matter, we move on to the body of the paper.
%% Sections are demarcated by \section and \subsection, respectively.
%% Observe the use of the LaTeX \label
%% command after the \subsection to give a symbolic KEY to the
%% subsection for cross-referencing in a \ref command.
%% You can use LaTeX's \ref and \label commands to keep track of
%% cross-references to sections, equations, tables, and figures.
%% That way, if you change the order of any elements, LaTeX will
%% automatically renumber them.
%%
%% We recommend that authors also use the natbib \citep
%% and \citet commands to identify citations.  The citations are
%% tied to the reference list via symbolic KEYs. The KEY corresponds
%% to the KEY in the \bibitem in the reference list below. 

\section{Introduction} \label{sec:intro}

Solar prominences are the plasma structures with high density and low temperature in the tenuous and hot corona \citep{Labrosse,Mackay10}.
The so-called   hedgerow type prominence is the quiescent prominence  that consists of long and tall blade-like palisades \citep{Engvold15}.
Observations have  shown that the plasma  within  such prominences typically exhibit both vertical and horizontal flows of the order of tens of km~s$^{-1}$ \citep{Engvold76,Chae10,Liu12}.
Similar to the bidirectional pattern of flows found in a filament \citep{Zirker98}, persistent horizontal flows that may reach long distances can be detected in    hedgerow prominences \citep{Chae08}. 
However, of particular interest are the thin, downward-directed mass flows with velocities much less than free-fall speeds.
The observational study by \citet{Liu12}  led the authors to suggest that the entire mass of a prominence could be swapped out on the order of a day by these vertical flows.
The mechanism responsible for then replenishing the drained prominence material has been suggested to be some form of coincident condensation process \citep[e.g.,][]{Antiochos,Keppens,Xia}  or a consequence of the draining material interacting with the transition region via the Rayleigh Taylor instability (RTI) \citep[see e.g.,][]{Keppens,Kaneko}. The RTI has also previously been invoked to explain the thin rising plumes that are typically coincident with the aforementioned falling structures \citep[e.g.,][]{Hillier}, not to be confused with the large \textit{voids} that have been observed to grow beneath prominences and are believed to be associated with flux emergence \citep[e.g.,][]{Berger10}.

The horizontal flows seen in the prominence are often interpreted as observational evidence that the plasma is supported  by the sagging of initially horizontal magnetic field lines \citep{Chae08,Shen15}.
 As such, it remains an open question as to how the knots observed to fall may do so through a medium permeated by horizontal magnetic field.
  \citet{Low12}  suggested that a downward resistive flow across the supporting horizontal field is a result of developing of a mass sheet singularity under the falling knots, while  \citet{Chae10} proposed instead that magnetic reconnection takes place around the descending knots.
However, \citet{Haerendel} suggested that the horizontal orientation of the external field is rather unfavorable for magnetic reconnection with the knots. 
 Assuming that  the plasma packet constituting the knots has a very high value of beta, the author suggested that the knots will behave like a diamagnetic body, which deforms the ambient field as it slips through it.
The downward acceleration is then counterbalanced by a friction force due to dissipation of Alfv{\'e}n waves along the horizontal magnetic field, and then  the downward motions keep almost  a constant speed.

 If the falling knots travel with constant downward speed along more vertical magnetic field, a hydrostatic pressure gradient  could provide an upward force which may balance  the force of gravity acting on the falling knots.
 Modeling the evolution of a density enhancement in a stratified atmosphere with uniform magnetic field, \citet{Mackay01}  found
that  this pressure gradient can build up   under the density enhancement,  causing it to fall  through the stratified atmosphere with speeds much less than the free-fall speed. 
Similarly, \citet{Oliver} investigated the dynamical evolution of a fully ionized plasma blob in an isothermal, vertically stratified corona.
The authors found that the presence of a heavy condensation  gave rise to the formation of a large pressure gradient that  opposed gravity, and eventually this pressure gradient became so large that the blob acceleration vanishes. 
In the model proposed by \citet{Low12}, the liberated energy is converted to dissipative energy that fuels the radiative loss of the prominence plasma. By contrast, \citet{Haerendel} suggested that the Alfv{\'e}n waves excited by the falling knots carry away the liberated energy.

Coronal heating is a topic dedicated to explaining how the corona may be heated up to a temperature of millions of degrees, far above that of the photosphere.
Alfv{\'e}n-wave turbulence is  a promising candidate to transport magneto-convective energy upwards along the Sun's magnetic field lines into the corona. 
\citet{McIntosh} detected Alfv{\'e}nic waves in  type II spicules, which could be  energetic enough to heat the quiet corona.
\citet{Samanta} showed the evidence that these spicules may originate from the solar surface. 
 However, some further investigations showed that  the wave energy communicated by spicules was in actual fact far less than that required to heat the corona of the quiet-Sun \citep{Thurgood,Weberg}.
As another candidate  for heating the corona, the nanoflare proposal suggested by \citet{Parker} involved the slow braiding of coronal field lines and the impulsive release of stored energy. 
Using  high-resolution coronal imager instrument onboard a rocket, \citet{Cirtain} suggested the magnetic braids in a coronal active region that are reconnecting, relaxing and dissipating sufficient energy to heat the
corona. However,  the braiding and relaxation are not immediately obvious as noted by   \citet{Zirker17}. The magnetic reconnection associated with the newly emergence flux  may also contribute to heat the quiet Sun \citep{Close,Zhang15}.

In this study, we have investigated the vertical motion of $H_{\alpha}$ knots within a solar prominence observed using the New Vacuum Solar Telescope \citep[NVST]{Liu01,Liu14}.
We have identified the rising and descending knots in the presented prominence,  before estimating  the velocity and acceleration of the knots on the plane of sky.
This allows us to investigate the  spatial distribution of temporal kinetics of the knots.
Furthermore, we have roughly estimated the loss rate of the gravitational potential from the descending knots and compared it with  heating power requirement for the corona in the quiet Sun.

   \section{Observation and Data analysis}
\subsection{Overview of the observation}
NVST is a ground-based multi-channel high resolution imaging system, including $H_{\alpha}$, G-band, TiO band and CaII 8542 \AA\  wavelengths. 
The studied prominence, located on the west limb of the Sun, was observed by NVST  from 02:20 to 07:10 UT on Jan 7, 2017.
During this period, NVST provides the $H_{\alpha}$ linecore image with  prefilter width of  0.25 \AA .
These images have a  temporal  cadence of 11s,  a spatial  sampling of $0\arcsec .16$, and a field of view of  $153\arcsec \times 153\arcsec$ .
A high resolution NVST image is normally reconstructed from at least 100 short exposure images.
 After reconstruction \citep{Xiang}, the actual resolution of an image of the chromosphere ($H_{\alpha}$, 6563 \AA) is better than $0\arcsec .3$. With such a spatial resolution, NVST can resolve many fundamental structures in the photosphere and chromosphere within a $3\arcmin $  field of view \citep{Xu}.

The observed prominence in  $H_{\alpha}$  image has a length of $\sim 80$ Mm and  an upper extent of $\sim 80$ Mm.
 Figure 1a presents an $H_{\alpha}$  image that is aligned to  the full-disk GONG  $H_{\alpha}$ image, which has
     a   spatial sampling of 	$1\arcsec .0$.

The Atmospheric Imaging Assembly \citep[AIA;][]{Lemen} on board the {\it Solar Dynamic
Observatory} \citep[{\it SDO};][]{Pesnell} provides  full-disk images of the corona with a  spatial sampling of $0\arcsec .6$.
Comparing the  $H_{\alpha}$  image with AIA/304 \AA\ image (Figure 1c), we can  see that the prominence observed in $H_{\alpha}$  is the northern part of a  larger  prominence   in 304 \AA\ image.

As shown on the bottom panels in Figure 1, the  images were rotated to ensue that the photospheric limb is horizontal. 
The AIA 211 \AA\ image (Figure 1f) that is taken at the same time with the NVST $H_{\alpha}$ image (Figure 1g) shows that a set of coronal loops appears under the prominence on the limb. 
Taken at an earlier time, the 211 \AA\ image presents that the coronal arcades (denoted by the arrow in Figure 1e) seem to be located around one  endpoint of the prominence (Figure 1e). 
Based on a high accuracy solar image registration procedure \citep{Feng,Yang14},
all of the NVST images are aligned to the same FOV as indicated by Figure 1g.
This will allow us to study the motions of the structures detected in the NVST images. 

\subsection{The technique of optical flow}

Time sequences of high-cadence images enable the identification and study of the dynamics features across the image plane.
The optical flow, referring to the  proper motion of a feature across the image plane,  may be used to determine the velocity field from two images.
The technique of local correlation tracking \citep[LCT;][]{November} and the differential affine velocity estimator   \citep[DAVE;][]{Schuck} are two kinds of  optical flow techniques that have been widely used in the solar research.

\citet{Thirion} presented a model to perform image-to-image matching by determining the optical flow between two images. 
The author illustrated the concept of this model by an analogy with Maxwell's demons,  as such this algorithm is usually named as non-rigid Demon algorithm.
\citet{Liu18} have examined the performance of three different methods (Demon, DAVE, and  LCT)  using  a photosphere and  chromosphere image provided by NVST.
After shifting the images 5.6 pixel in x direction and 0.2 pixel in y direction,  they noted that the Demon found displacements are more close to  the \textit{true}  displacements and then they suggested that the Demon  algorithm outperforms traditional LCT and DAVE methods in estimating the  analog displacements both smaller and  bigger than   one pixel.  

Here, we used both the DAVE and Demon algorithms to obtain the optical flow between two NVST $H_{\alpha}$ images.
 The earlier image is set as the  reference image and  and the later image as the target image.
As an example, Figure 2a shows a reference image and Figure 2d shows the difference image between the reference and target image.
 The DAVE and Demon velocity fields are indicated by arrows on Figure 2b and 2c , respectively.
To test the performance of each technique, we warped the reference image to align with the target image based on the resulted optical flow.
Ideally, the warped image would be identical to the target image.
Figure 2e (2f) presents the difference image between the target image and the warped image based on DAVE (Demon).
Comparing the warping image based on DAVE, the Demon-based warping image has less difference from the target image.
%Further work is in  preparation to test the performance of Demon and other optical flow in the various solar images.
 Based on these results, the Demon-based optical flow is  chosen to study the velocity of the knots in the prominence.
As shown in the Appendix,  the errors of the velocities measured by Demon is about 2 km~s$^{-1}$.

\subsection{Identifying and tracking knots}
Different from the vertical thread in the prominence, the bright knots appear to be brighter and shorter in the vertical direction.
Here, we  used morphological approach to identify the rising or descending knots in the prominence.
Firstly,  since the bright knots are located on  nonuniform background,  contrast of the $H_{\alpha}$ image was enhanced using contrast-limited adaptive histogram equalization \citep[CLAHE;][]{Zuiderveld}.
      Figure 3a presents one of the enhanced $H_{\alpha}$ images. A representative bright knot is denoted by the arrow in Figure 3a.
Secondly, to further enhance contrast of the bright knots on the images, we performed morphological top-hat filtering on the enhanced $H_{\alpha}$ image, the bright knots were then separated from the background  (Figure 3b).
Based on a threshold of intensity value on the top-hat transformed image, the bright knots can be identified, as the color ( magenta and green) contours outlined (Figure 3b).
Third,  
to distinguish between the rising and falling blobs, and other uninteresting features, we selected the structures that have a vertical displacement larger than its initial vertical extent.
As shown in Figure 3b, the  magenta contours  denote the rising or descending bright knots that are selected, while the  uninteresting features are colored green.
Each point in Figure 3c  denotes the   location where each knot starts to be identified. The knots with rising and descending motions are colored red and blue, respectively.
It is obvious from the distribution of points that the rising knots are often located in the higher altitude, while the descending motions in the low altitude.
  In Figure 3d, the identified knots are colored based on the vertical velocity estimated from Demon.

\section{Results}

 Applying the algorithm outlined above to the $H_{\alpha}$ images,  we identified and tracked 6194 moving bright knots.
 The widths of the knots range from 1 Mm to 3 Mm (Figure 4a). 
The heights of the knots to solar surface range from  2\,--\,85~Mm (Figure 4b). 
 Among these tracked knots, 4655 knots show downward motions, and the remaining 1539 knots show upward motions. 
The knots travel upward or downward with a vertical distance of several Mm (Figure 4c). 
Figure 4c shows the distribution of the vertical distances that the knots traveled. Here, 
positive and negative value means that the knot has experienced a downward and upward motion, respectively.
We found that the  majority  of knots disappeared from the $H_{\alpha}$ image after rising or falling less than 4 Mm. These knots are visible  in less than 10   $H_{\alpha}$ images and their lifetimes range from 1 to 2 minutes. However, 713 of the 4655 downward knots were detected to travel longer than 4 Mm in the vertical direction, among which 75 knots descend more than 10 Mm.

The velocity of a knot is estimated by averaging the Demon velocities of the knot detected at different times.
  The histogram in Figure 4d shows  the vertical velocity distribution for the identified knots.
 Again, more knots are found to display downward velocity.  The descending knots fall at the speed of about 18 km~s$^{-1}$, while the other knots rise at the lower speed of  10 km~s$^{-1}$.
 
  The acceleration  $a$ of a knots is obtained by    a least square linear fit for the velocity, such as $V_{fit,i}=a v_{i}+b$, where $v_{i}$ is the vertical velocity of a knots observed in the ith moment. The histogram of $a$ (Figure 4e) shows that    the distribution of knot acceleration peaks at a value of zero with a FWHM of only 0.42$g_{Sun}$, where $g_{Sun}$=0.272 km~s$^{-2}$ near the solar surface.

  Figure 4g presents the histogram of the brightness of knots. Again, the brightness of a knot is defined as the average of that of the  knot detected at different moments.
  We defined the change rate of the brightness of an identified knot as $\Delta B=(B_{fit,N}-B_{fit,1})/\bar B_{fit}$, where $B_{fit,i}=a b_{i}+b$ is a linear fit to the brightness  $b_{i}$ of a knot  detected at the ith image, i=1, 2, ... N, and a knot is identified in N images.
The $\Delta B$ greater than zero means that the total brightness increases when a knot rises or descends, while $\Delta B$ lower than zero indicates  the brightness decreases.
The histogram in Figure 4h shows that the distribution of $\Delta B$  is close to a gaussian profile with a peak at -0.10, indicating that 
the number of knots whose brightness decreases over time is slightly bigger.

It is worth noting that, however, the knots  disappeared in the  $H_{\alpha}$  image after the knots were traced in several or more images.
 In order to isolate the final velocity and the brightness of the knots that  were recognized in the  $H_{\alpha}$  images,
 we defined 
 \begin{equation}
 \Delta v_{end}=\frac{v_{end}-V_{fit, end}}{V_{fit, end}},  
  \Delta b_{end}=\frac{b_{end}-B_{fit, end}}{B_{fit, end}}
   \end{equation}
   where the term ``end'' denotes the image in which a knot was recognized at last time.
   The larger value of $\Delta v_{end}$ ($\Delta b_{end}$) indicates larger change in the vertical velocity (total brightness) of a knot before it disappears.
   The histogram of $\Delta v_{end}$  (Figure 4f) shows a  gaussian profile with peak at -0.1, while the histogram of $\Delta b_{end}$ (Figure 4i) fits a Gaussian distribution with the peak at -0.39.
   Therefore, at a time when these knots were about to disappear on the $H_{\alpha}$ image,  the brightness of knots  decreased impulsively but their descending velocity did not change significantly. It implies that  the knots may keep falling after they cannot be identified in the $H_{\alpha}$ images.

   Figure 5a shows an averaged $H_{\alpha}$ image, in which each pixel is an average of the brightness of the knots that are detected  at the same location at different times.
    Similarly, Figure 5b and 5c shows the averaged Demon velocity  distribution of the knots in the horizontal and vertical direction, respectively.
    Figure 5b shows that the horizontal velocity is close to  30 km~s$^{-1}$ on the bottom of the prominence and is greater than  5 km~s$^{-1}$ in most areas at higher altitude.
   In Figure 5c, the upward vertical velocity  could be found predominantly in the upper portions of the prominence whereas the downward velocity were present at seemingly all altitudes with the strongest signatures at the bottom of the prominence.
   
   To  detail the value of the speed of the falling knots, Figure 5e shows the distribution of the vertical velocity in the  downward direction.
 In the area denoted  by  the box  in Figure 5e, the downward velocity of the knots is greater than  25 km~s$^{-1}$.  Comparing to the FOV of AIA 211 \AA\ , the  box covers a set of coronal loops (Figure 5d). It is thus possible that the quickly falling knots may run along magnetic field that is marked by the coronal loops. 
 The  majority  of knots outside the box were found to descend with  a velocity lower than 25 km~s$^{-1}$.
    Overall, the largest downward velocities were located at the bottom of the prominence.
  
 The acceleration of a knot  was estimated from the result of linear fit to the   vertical speed of an identified knot in three moments.  
 The map in Figure 5f then presents   averaged acceleration of the identified knots, in which the  dashed-line denotes the height of 7 Mm above the solar surface. The  majority of the knots at a height above 7 Mm  appear to have had a downward acceleration much less than the gravitational acceleration experienced  at the solar surface.
 However, within about 7 Mm above solar surface, the knots underwent a  rapid deceleration, with the value of acceleration as high as $-g_{sun}$ .

  \section{Discussions}
  In this study,we have identified and tracked the knots   within a solar prominence that was observed by the NVST.
   Although knots were observed to have propagated both upwards and downwards (with respect to the solar surface), a larger percentage of the motions were recorded in the downward direction. These upward motions may be a consequence of the initially downward-directed motions \citep[see the RTI as explored by e.g.,][]{Keppens}, and we may in-turn speculate that this explains their prevalence at the upper levels within the prominence. However, in the absence of additional information we continue below with a focus on the more common downward motions.

  By identifying  and tracking the descending knots in the prominence, we found that the knots reached a higher speed in the low altitude.
  The knots that reached the maximum speed of 30 km~s$^{-1}$ were co-spatital with a set of coronal loops observed in AIA 211 \AA , thereby it is possible that the knots with high speed fell along the magnetic field lines denoted by the observed loops. If so, the gravity acted on the knots may be balanced by the pressure gradient when the knots  descended along the magnetic field \citep{Mackay01}. 
However, as discussed below,  this picture may not  sufficiently explain the motions of the knots such as those  with lower downward speed  at higher altitudes.

Falling through the  stratified atmosphere threaded with a vertical magnetic field, a knot is expected to achieve greater speed if their density is higher or  they fall in a rarer environment  \citep{Oliver}.
Assume that  each knot is optically thin in $H_{\alpha}$, the higher density of a knot in the same altitude would be brighter in $H_{\alpha}$ images \citep{Heinzel15}. 
   However, no evidence  showed that the knots having higher descending speed had the  higher density, because that the brightness of the knots  did not have the correlation to the vertical velocity of the knots.
   On the other hand, vertically stratified corona  has the vertical scale height of   about 120 Mm. 
   It means that,  with the height of 80 Mm, the density of the corona   in the high altitude is significant rarer than in the low altitude.
   Therefore, it is expected that the knots  in the high altitude could reach higher downward speed if all of the knots fall along the  vertical magnetic field in the  stratified corona, which is contradictory to the observation presented here.
    Instead, we are inclined to view  that most of the knots  fell across more horizontal field as suggested in \citet{Mackay01}.  This is particularly important for those knots that were not co-located with the loops viewed in AIA 211 \AA\ .
   
   The descending knots  were measured to have fallen at speeds of approximately 18 km~s$^{-1}$, which is consistent with previously-established observational values of 10\,--\,15 km~s$^{-1}$ \citep{Berger08,Chae10,Li18}.
   Using a three-dimensional magnetohydrodynamic simulation including optically thin radiative cooling and nonlinear anisotropic thermal conduction, \citet{Kaneko} found that the downward speed of the dense plasma is approximately 12 km~s$^{-1}$.
   In the prominence studied here, a few knots fell at speed as high as about 40 km~s$^{-1}$.
  \citet{Freed}  
  reported on vertical plasma motions within prominences that reached magnitudes up to 30 km~s$^{-1}$, although the authors also noted that the FLCT method systematically underestimates the \textit{true} velocity.
   MHD modeling of prominence dynamics also  indicated downward motions of dense plasma at up to 60 km~s$^{-1}$  prevailed  instantaneously \citep{Keppens}

   Since the knots  displayed an approximately constant downward motions,  most of the liberated gravitational potential energy may not  have manifested itself as an increase in kinetic energy.
As suggested by \citet{Low12}, if the knots flow downward across the supporting magnetic field, the liberated gravitational potential energy could be conserved into dissipative energy that fuels the radiative loss of the prominence plasma.
   Based on the model proposed by \citet{Haerendel}, the descending knots would excite the Alfv{\'e}n wave propagating along the horizontal magnetic field. It means that the gravitational potential energy could be converted into the wave energy, which may  contribute to the heating of the corona in the quiet Sun.

   The loss of the gravitational potential energy is equal to $Mg_{Sun}h$, where $g_{sun}$ is the gravitational acceleration of 0.272 km~s$^{-2}$ near the solar surface.    In the present study, 
    we  took $h$ as  the distance from which an identified knot  had fallen. 
   The mass of a cool prominence is roughly equal to $M \approx 1.4 N_{H} m_{H} V $ \citep{Heinzel96},  where $N_{H}$ and $m_{H}$ is the mean neutral hydrogen number density and the mass of the hydrogen atom, respectively, and $V$ is the volume occupied by the prominence plasma.
   
The literature values of the electron density within prominence varied greatly and within a range of   $10^{9}$ to $10^{11} ~cm^{-3}$ \citep{Labrosse}. Some of these variations were, no doubt, due to differences between the various techniques that are used, but there  were likely to also be variations between parts of the same prominence.
Unfortunately, few studies have focussed specifically on the density of the knots observed within solar prominences.
 Assuming that the knots have the high-beta magnetic structure, \citet{Haerendel}   estimated that the peak density of the knots must be as high as $3 \times 10^{12} ~cm^{-3}$ with an average density of $5 \times 10^{11} ~cm^{-3}$.
Accordingly, we take the density of the knots as $5 \times 10^{11} ~cm^{-3}$.
To determine the volume of the knots, we assume that  the length of the knots in the line of sight has the same scale with that in the plane of sky.
Based on the above assumptions,  the loss of gravity potential energy from the detected falling knots is about $5.3 \times 10^{21} erg$ during the period of $\sim 5$ hours. 
This is equivalent to a potential energy loss rate of  $3 \times 10^{17} erg ~s^{-1}$ for the falling knots in this prominence.

 We assume that  the  majority of  the knots were not falling through additional prominence magnetic field, but instead across the horizontal field  of the  background corona. 
 This  implies that  the converted gravitational energy is not contributing to the heating within the prominence itself. 
  Under this assumption, it is then possible that the energy released from the falling knots may be radiated away from these knots and into the surrounding environment \citep[as in][]{Haerendel}. In the quiet-Sun, the heating power requirement is about $10^5~ erg~cm^{-2} ~s^{-1}$ \citep{Withbroe}. If we roughly assume that the area of quiet Sun is equivalent to the surface area of the whole Sun, the gravitational energy lost by these falling knots amounts to 1/2000 of the energy needed to heat the quiet-Sun.

It is worth noting that the individual knots may continue to fall  after they disappeared   according to the $H_{\alpha}$  linecore image.
This is supported by the fact that the knots did not  display any significant decrease in the falling speed before they disappeared in our $H_{\alpha}$  image.
The linewidth of NVST $H_{\alpha}$  prefilter is 0.25 \AA\, which means that the knots disappear on the $H_{\alpha}$ images when the LOS speed of the knots exceeds 5 km~s$^{-1}$.
During the 5 hours of observations, the horizontal velocity of knots on the image plane was often more than 5 km~s$^{-1}$. 
 Assuming that the LOS velocity may reach the same order of magnitude as the horizontal velocity  on the plane of the $H_{\alpha}$ images, we suggest   that a significant portion of the knots may experience further downward motion not captured in the $H_{\alpha}$ observations.
Moreover, it is also worth drawing attention to the differing behaviour of the knots above and below a height of 7 Mm. Below 7 Mm it was clear in Section 3 that the knots experienced strong decelerations as they approached the surface, however those above 7 Mm appear to have maintained a constant velocity.
 Therefore, considering instead that these knots may have traversed a significant height, they  could release  gravity potential energy of  the order $3.1 \times 10^{22} erg$, which is equivalent to a potential energy loss rate of  $1.9 \times 10^{18} erg ~s^{-1}$.
 Hence, the energy loss rate from this studied prominence could  increase to  1/320 of the heating power requirement for the corona in the quiet Sun.
 Of course, this can be viewed as an ideal upper limit for the loss rate of gravitational potential energy from the knots in the prominences, and corresponding conversion to dissipated thermal energy.
Indeed, the observation from off-band $H_{\alpha}$ wavelengths will allow us to further constrain this conclusion with an ability to detect the descending knots that have LOS speed greater than 5 km~s$^{-1}$.

 The prominences are not ubiquitous structures and are therefore unlikely to be responsible for heating the entire quiet-Sun. 
However, the prominences  same as or   larger than one present here are often observed in the solar limb.
Moreover,  only the prominence whose spine is almost parallel to the plane of sky could show its details on the image.
When the spine of the prominence runs along the LOS, the prominence appears as a very narrow structure on the edge of the Sun.
Furthermore, there are  small prominences which may not be clearly observed.
 Of course, prominences  often appear as filaments on the solar disk and a lot of filaments with various scales can be observed simultaneously. Taken together, it is likely that a considerable number of prominence exists in the solar corona at the same time.
Therefore, within the bounds of our assumptions, the possibility of ubiquitous, falling knots may well contribute in a non-negligible way to the local heating of the surrounding corona.

 \section{Conclusion}
We have investigated the descending knots in a solar prominence observed by NVST.
Our major findings and their implications are as follows.
\begin{enumerate}
\item Using the Demon method, we found that most of the detected knots fell with an approximately constant speed.
\item By analysing the distributions of the downward velocities for these falling knots, we suggested that  the constant velocity is consistent with the theory of plasma packets slipping across horizontal magnetic field.
\item Assuming that the  gravity potential energy lost by these detected falling knot could convert to the thermal energy of the corona, we speculated that released potential energy by the falling knots could  heat the surrounding corona.
\end{enumerate}
A robust estimation of the density of the knots in the prominence will allow us to precisely estimate the loss rate of the gravitational potential of the falling plasma.  
 Additional modelling work is also required in order to reveal how the interaction between the falling dense plasma and the magnetic field generates the  Alfv{\'e}n wave, dissipation of which is capable of contributing to the heating of the local corona.

    \section{Acknowledgements}
The authors sincerely thank the anonymous referee for detailed comments and useful suggestions for improving this manuscript. 
%The edits and rewordings by referee have provided explanations in a more clear and concise manner.
    Y. B. is supported by the  Natural Science Foundation of China under grants 11873088 and U1831210, and Young Elite Scientists Sponsorship Program  by CAST.
    T. L.  is supported by the National Key Research and Development Program of China (2019YFA0405000).
    This work is also  supported by the Natural Science Foundation of China under grants  11633008,   11773072, 11873027,  11933009, 11773039, 11933009,  11703084, and 11763004, CAS grant QYZDJ-SSW-SLH012, the grant associated with the project of the Group for Innovation of Yunnan Province, and the project supported by
     the Open Research
Program of the Key Laboratory of Solar Activity of Chinese
Academy of Sciences (grant No. KLSA201710). 
This work used the DAVE/DAVE4VM codes developed by the Naval Research Laboratory. The NASA/SDO and GONG data used here are courtesy of the  AIA   and GONG science teams.

\section{Appendix}
\subsection{Uncertainty of DAVE and Demon velocities}

In order to ascertain the uncertainty in the DAVE and Demon velocity, an experiment was performed on synthetic data sets.
We created a synthetic image by shifting one of the $H_{\alpha}$ images a random number of pixel away from its original
position in both the horizontal and vertical direction.
The random number was chosen to range from 0.5 pixel to 4.5 pixel, since the velocities of detected knots range from 5 km~s$^{-1}$ to 50 km~s$^{-1}$.
Between the original and the shifted image, all of the knots were moved at a known distance.  
The procedure was repeated  a total of  200 times.
Accordingly, we could compare the 200 sets of known displacement  of each knot and the corresponding  DAVE or Demon found displacement.
If the found displacements are equal to the known displacements, the points should scatter around the x=y line in the scatter diagram (Figure 6).
For the most of knots, the DAVE found displacements were lower than the known displacements (Figure 6a), indicating that
the DAVE method often underestimates the velocities of these knots.
However, the differences between the Demon found displacements and the known displacement were always lower than 0.18 pixel  (Figure 6b). This implies that the errors of the velocities  measured by Demon amounts to 2 km~s$^{-1}$.
We then suggest that the Demon method could provide better estimated velocities of the knots that are detected in the $H_{\alpha}$ image.

\begin{figure}[h]
\includegraphics[scale=0.25]{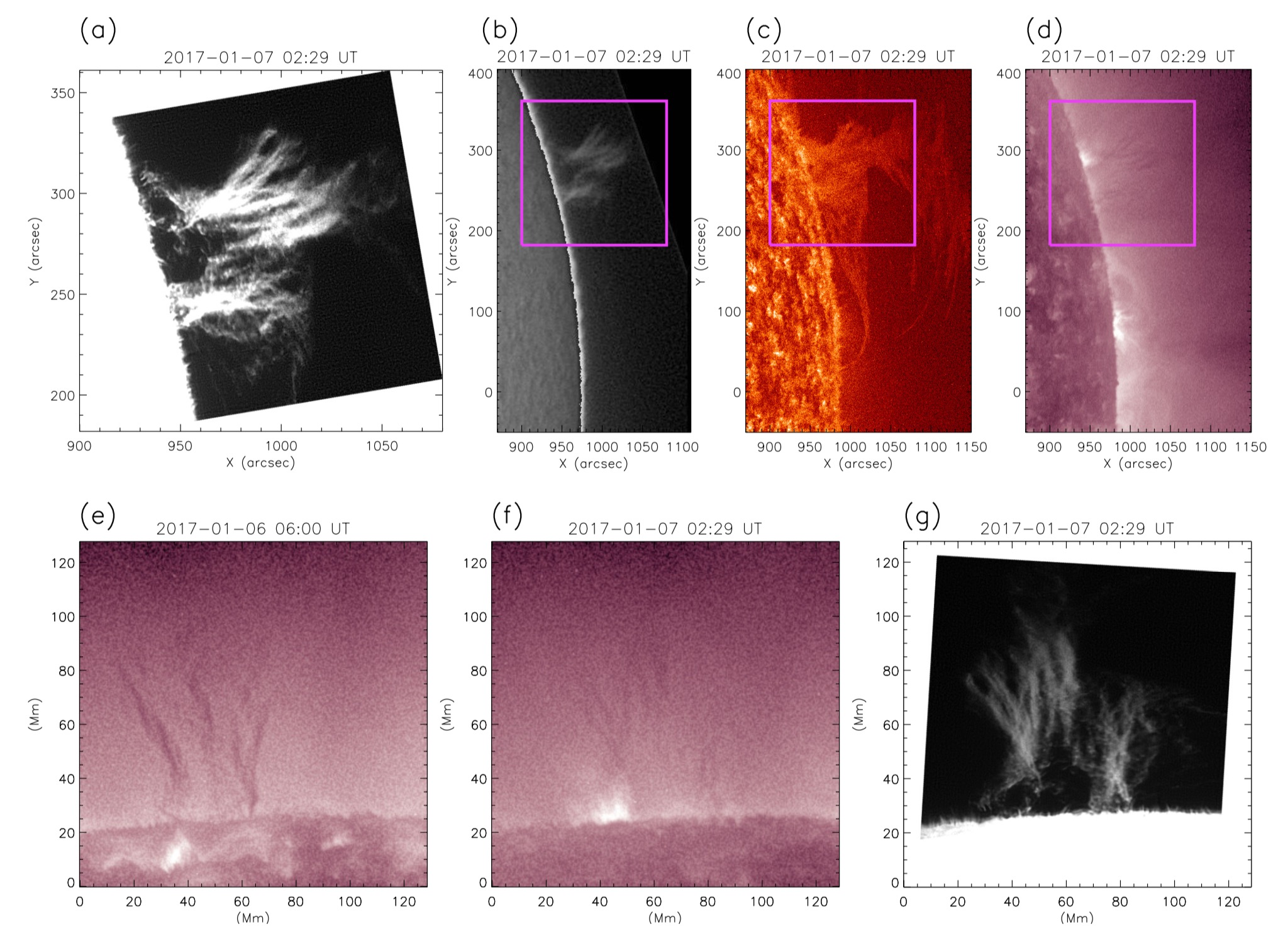}
\caption{
(a, g) NVST $H_{\alpha}$ image.
 (b) GONG $H_{\alpha}$ image.
 (c) AIA 304 \AA\ image.
 (d,e, f) AIA 211 \AA\ image.
The pink  boxes on the three top panels indicate the position and FOV of panel (a) , which is the full view of the NVST.
The   images on the bottom panels  (e, f,  g)  are rotated so that the photospheric limb is horizontal.
Each AIA image was enhanced by using the Multiscale Gaussian Normalization(MGN) technique \citep{Morgan}, which  is very effective at revealing faint fine-scale details of the prominence studied here. 
 }
\end{figure}

\begin{figure}[h]
\includegraphics[scale=0.5]{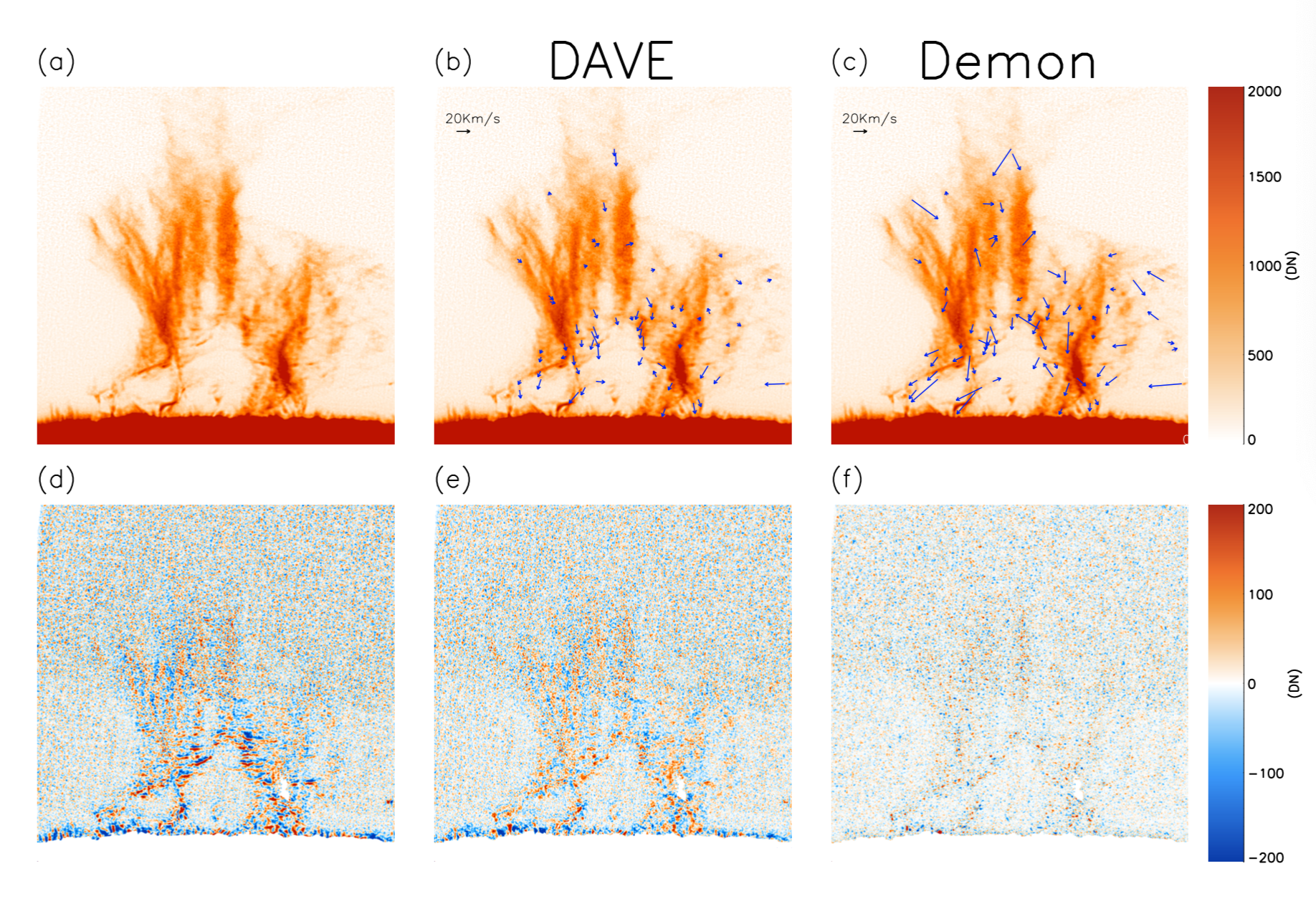}
\caption{
(a, b, c) NVST $H_{\alpha}$ images taken at 02:22:29 UT.
(d) The difference image between the reference and target image, which refers to NVST  $H_{\alpha}$ image taken at 02:22:29 UT and 02:22:51 UT, respectively.
On panels (b) and (c), the arrows indicate the optical flow based on technique of DAVE and Demon, respectively.
(e, f) The difference images between the reference and warping image. 
In (e) and {f}, the warping image refers to the reference images that is warped to align with the target image based on the optical flow of DAVE and Demon, respectively.
 }
\end{figure}

\begin{figure}[h]
\includegraphics[scale=0.25]{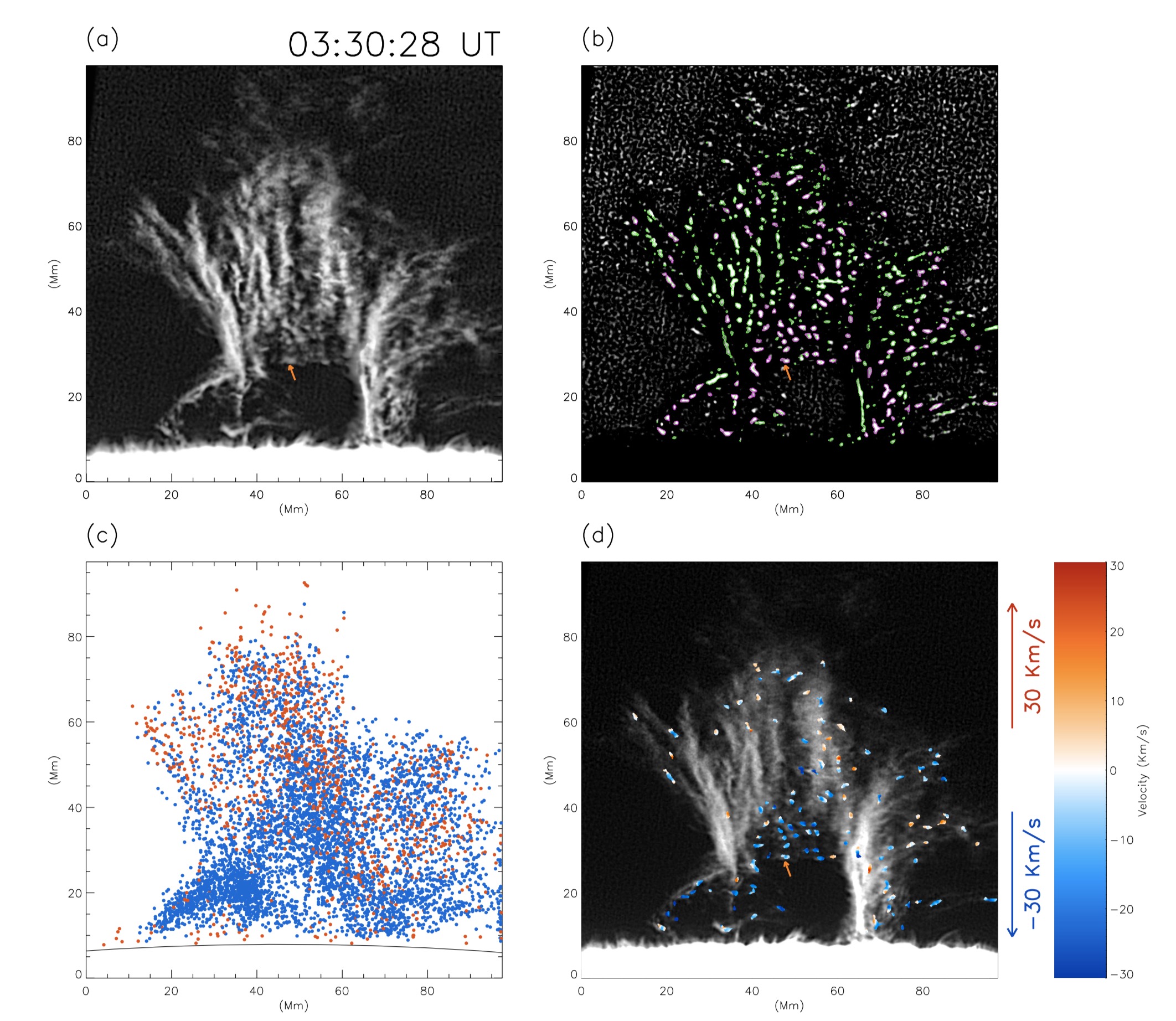}
\caption{
(a) NVST $H_{\alpha}$ image that is CLAHE enhanced.
(b) The identified bright knots are outlined with color contours. The mega contours refer to the knots that have the value of vertical displacement lager than the height of the knots themself. The green contours denote the left knots.
(c) Each point denotes the   location where each knot starts to be identified. The red and blue points indicate the  knots that show rising and descending motions, respectively.
\textbf{The dark line denotes the solar limb.}
(d) The identified knots are colored based on the value of vertical Demon velocity. 
On each panel, the orange arrow denotes a knot as an example.
 }
\end{figure}

\begin{figure}[!h]
\includegraphics[scale=0.6]{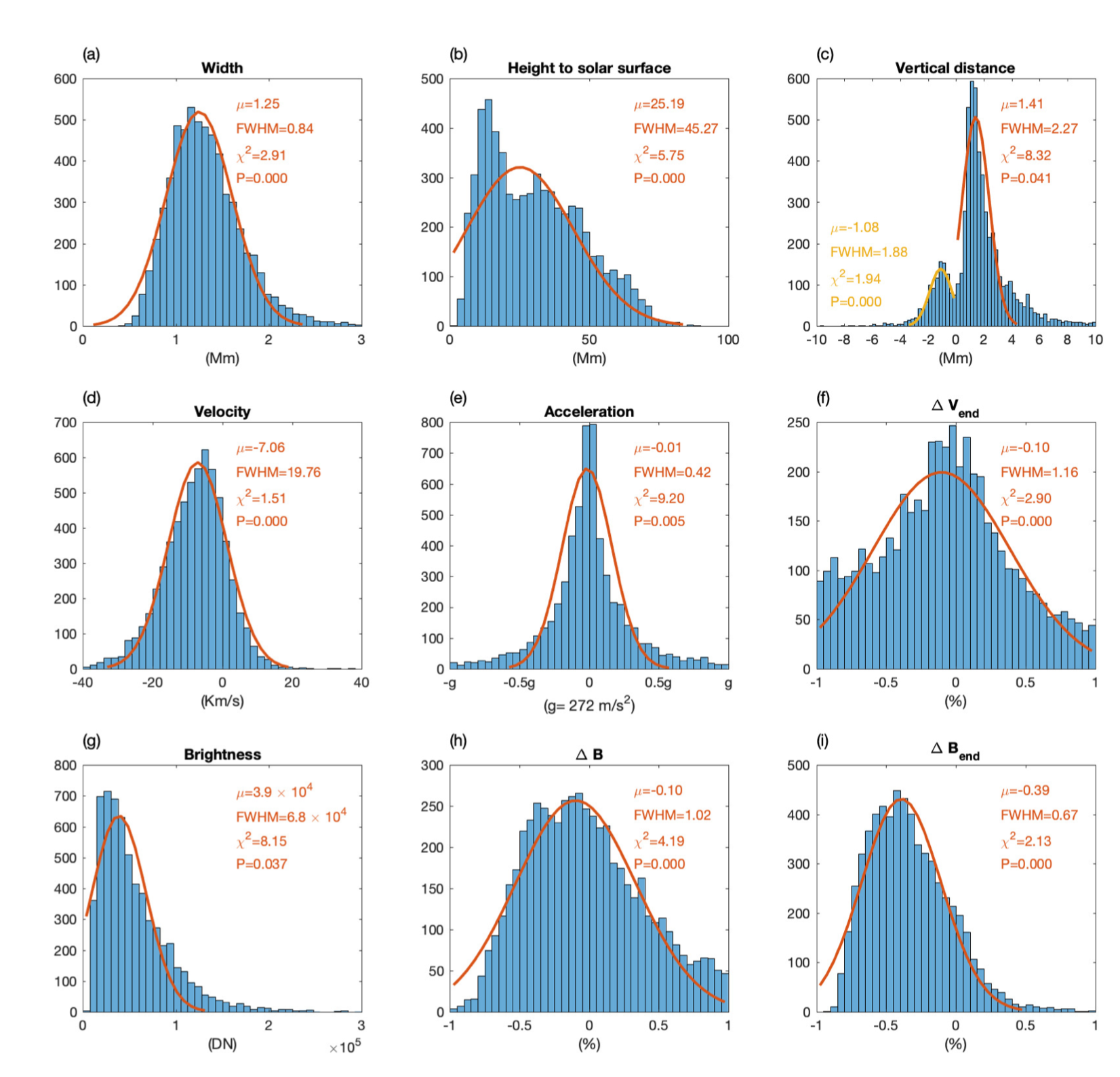}
\caption{
Histograms showing the distributions of width (a), the height to the solar surface  (b), vertical distance of fall (c), vertical velocity (d), vertical acceleration (e), $\Delta V_{end}$ (f), Brightness (g), $\Delta B$ (h), and $\Delta B_{end}$  (i) of the identified knots, respectively. 
On each panel, the  line represents the central part of a Gaussian function $\mathcal{N}(\mu,\sigma^2)$ fitting the main part of binned data. 
The fitting parameters are shown in each panel, including the parameter $\mu$, full width at half maxima (FWHM) of $\sim 2.35 \sigma$,  chi-square ($\chi^2$), and \textit{P}-value of each fitting.
In general,
we regard the Gaussian  function is a suitable fit to the data if its \textit{P}-value
is smaller than 0.05. Accordingly, the Gaussian shape is a reasonable assumption  for the  main part of each distribution.
 }
\end{figure}

\begin{figure}[!h]
\includegraphics[scale=0.5]{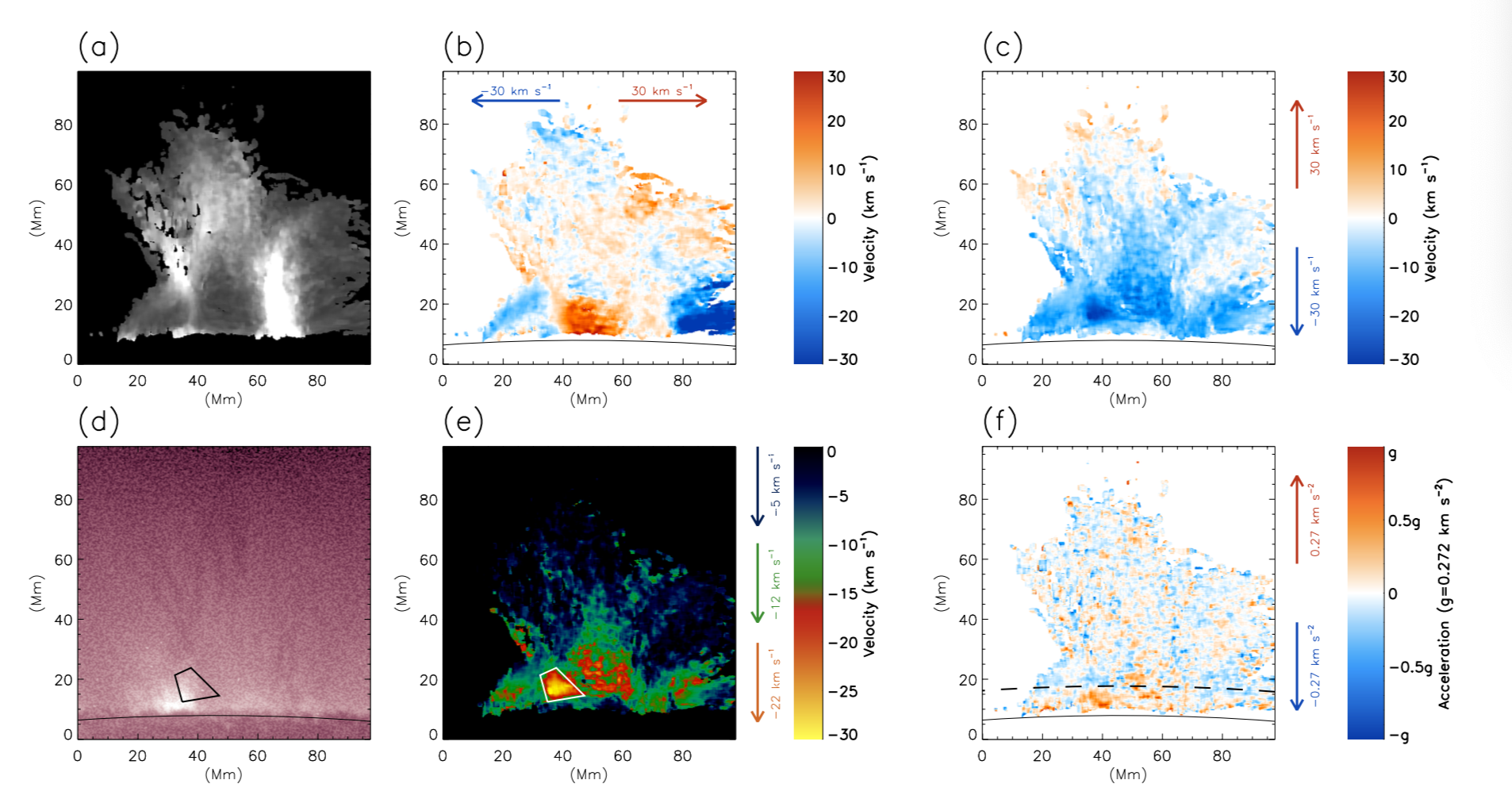}
\caption{
Maps of averaged brightness (a), vertical velocity (b), horizontal velocity (c), downward velocity (e), and vertical acceleration (f) of the  identified knots.
 On these maps, the value in each pixel is an average of the values  in each pixel at different times.
 (d) AIA/211 \AA\ image. 
 On panel f, the dashed line indicates a line with height of 7 Mm above solar surface.
 }
\end{figure}

\begin{figure}[!h]
\includegraphics[scale=1]{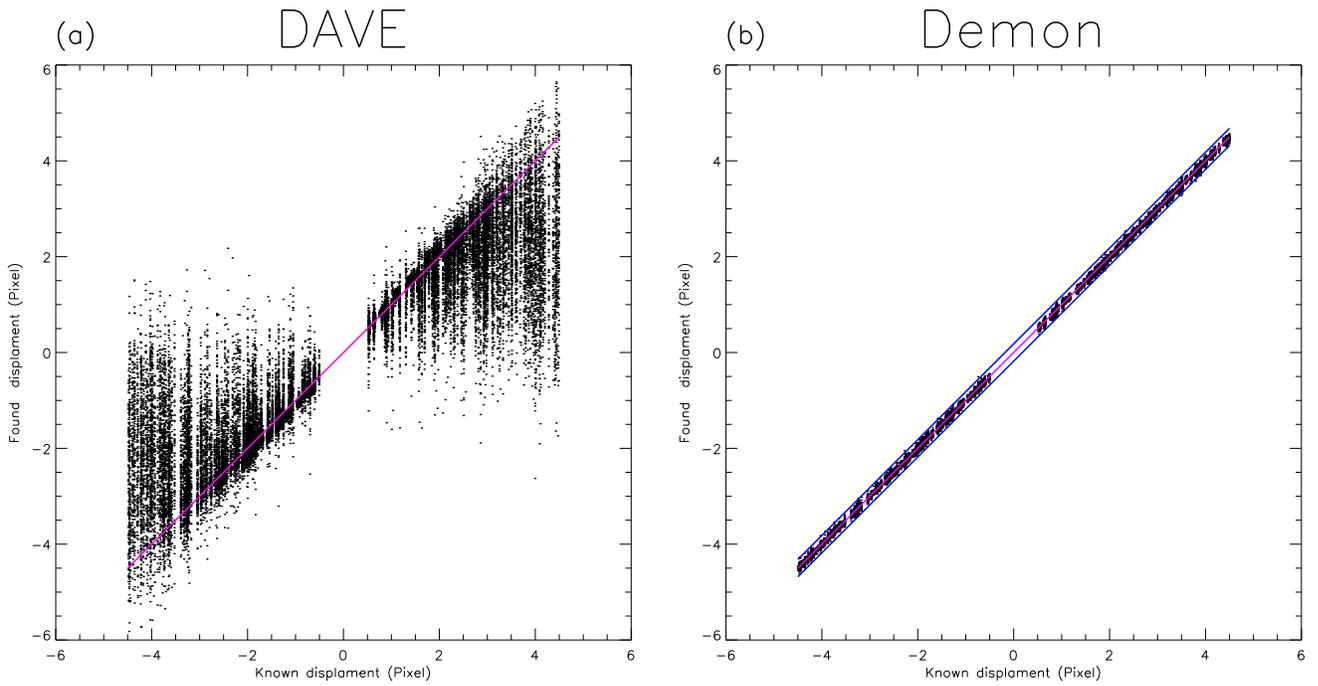}
\caption{
A plot of the  known velocities and the found velocities of each detected knots. On Panel   (a) and (b),  the velocities are  found based on the method of  DAVE and Demon, respectively.
On each  panel, the magenta line is a  x=y  line  passing through the origin.  On  panel (b),  the blue lines are parallel to and have a space of 0.18 pixel with the magenta line.
 }
\end{figure}

\end{document}